\documentclass[twocolumn]{emulateapj}
\usepackage{apjfonts}

\usepackage{natbib}
\usepackage{epsfig}
\usepackage{rotating}
\usepackage{hyperref}

\newcommand{\be}{\begin{equation}}
\newcommand{\ee}{\end{equation}}

\newcommand{\eg}{\emph{e.g.}}
\newcommand{\kms}{\mbox{km\,\ensuremath{\rm{s}^{-1}}}}
\newcommand{\ebv}{$E_{B-V}$}
\newcommand{\dib}{$\lambda$}
\newcommand{\cp}{C$_{60}^+$}

\newcommand{\espadons}{ESPaDOnS}
\newcommand{\ki}{K\,{\sc i}}

\submitted{ApJL, accepted for publication April 2019}

\shortauthors{Cordiner et al.}

\begin{document}

\title{Confirming interstellar C$_{60}^+$ using the Hubble Space Telescope}

\author{M. A. Cordiner\altaffilmark{1,2}, H. Linnartz\altaffilmark{3}, N. L. J. Cox\altaffilmark{4},  J. Cami\altaffilmark{5}, F. Najarro\altaffilmark{6}, C. R. Proffitt\altaffilmark{7}, R. Lallement\altaffilmark{8}, P. Ehrenfreund\altaffilmark{3,9}, B. H. Foing\altaffilmark{10}, T. R. Gull\altaffilmark{1}, P. J. Sarre\altaffilmark{11},  S. B. Charnley\altaffilmark{1}}


\altaffiltext{1}{NASA Goddard Space Flight Center, 8800 Greenbelt Road, Greenbelt, MD 20771, USA}
\email{martin.cordiner@nasa.gov}
\altaffiltext{2}{Department of Physics, Catholic University of America, Washington, DC 20064, USA}
\altaffiltext{3}{Sackler Laboratory for Astrophysics, Leiden Observatory, Leiden University, PO Box 9513, NL 2300 RA Leiden, Netherlands}
\altaffiltext{4}{ACRI-ST, 260 route du Mon Pintard, Sophia Antipolis, France}
\altaffiltext{5}{Department of Physics and Astronomy and Centre for Planetary Science and Exploration (CPSX), The University of Western Ontario, London, ON N6A 3K7, Canada}
\altaffiltext{6}{Departamento de  Astrof\'{\i}sica, Centro de Astrobiolog\'{\i}a (CSIC-INTA), ctra. de Ajalvir km. 4, 28850 Torrej\'on de Ardoz, Madrid, Spain}
\altaffiltext{7}{Space Telescope Science Institute, 3700 San Martin Drive, Baltimore, MD 21218, USA}
\altaffiltext{8}{GEPI, UMR8111, Observatoire de Paris, 5 Place Jules Janssen, 92195, Meudon, France}
\altaffiltext{9}{George Washington University, Washington DC, USA}
\altaffiltext{10}{ESA ESTEC SCI-S, Noordwijk, The Netherlands}
\altaffiltext{11}{School of Chemistry, The University of Nottingham, University Park, Nottingham, NG7 2RD, UK}

\begin{abstract}
Recent advances in laboratory spectroscopy lead to the claim of ionized Buckminsterfullerene (\cp) as the carrier of two diffuse interstellar bands (DIBs) in the near-infrared.  However, irrefutable identification of interstellar \cp\ requires a match between the wavelengths and the expected strengths of all absorption features detectable in the laboratory and in space. Here we present Hubble Space Telescope (HST) spectra of the region covering the \cp\ 9348, 9365, 9428 and 9577~\AA\ absorption bands toward seven heavily-reddened stars. We focus in particular on searching for the weaker laboratory \cp\ bands, the very presence of which has been a matter for recent debate. Using the novel STIS-scanning technique to obtain ultra-high signal-to-noise spectra without contamination from telluric absorption that afflicted previous ground-based observations, we obtained reliable detections of the (weak) 9365, 9428~\AA\ and (strong) 9577~\AA\ \cp\ bands. The band wavelengths and strength ratios are sufficiently similar to those determined in the latest laboratory experiments that we consider this the first robust identification of the 9428~\AA\ band, and a conclusive confirmation of interstellar C$_{60}^+$.
\end{abstract}


\section{Introduction}

The diffuse interstellar bands (DIBs) are a series of several hundred broad absorption features that occur in optical-NIR spectra of stars as their light passes through the diffuse interstellar medium (ISM). The total DIB absorption cross-section indicates the presence of a large quantity of interstellar material, much of which is believed to be carbonaceous in nature \citep{cor11}, but the identities of the DIB carriers have been elusive despite dedicated observational, laboratory and theoretical efforts since the early 20th century \citep{her95,sar06,cam14}. Identifying the DIB carriers will have important implications for our understanding of the ISM and its constituents. DIBs have recently been used as tracers of small-scale interstellar structure \citep{cor13}, and DIB strength variations can reveal the interaction between supernovae and their surrounding environments \citep{mil14}. Due to their ubiquity in local and distant galaxies, DIBs are potentially invaluable probes of interstellar chemistry and physics throughout the universe \citep{cor14}.

Based on photofragmentation spectroscopy of C$_{60}^+$--He complexes at {very} low temperature, \citet{cam15} claimed identification of a pair of diffuse interstellar bands in the NIR with C$_{60}^+$.  This work came almost two decades after the initial association of the 9577~\AA\ and 9632~\AA\ DIBs with this molecule \citep{foi94,foi97}. As a complement to the prior observations of fullerenes in circumstellar and nebular environments \citep{cam10,sel10,ber13}, the ongoing work on C$_{60}^+$ in the diffuse ISM is causing a significant shift in our understanding of the possible inventory of very large molecules in low-density interstellar clouds \citep[\eg][]{omo16} --- the largest molecules definitively detected to-date in the diffuse ISM have only three atoms heavier than hydrogen \citep{sch14,lis18}.

As discussed by \citet{gal17,gal17b,cor17,lal18}, the case for interstellar C$_{60}^+$ has not yet been proven beyond doubt. The laboratory studies of \citet{cam16,kuh16,spi17} have identified the presence of five near-infrared \cp\ absorption features at 9348.4, 9365.2, 9427.8, 9577.0 and 9632.1~\AA, with recently updated (peak) cross section ratios 0.09:0.26:0.17:1.0:0.84 \citep{cam18}. So far, the detection of interstellar C$_{60}^+$ has been grounded in the wavelength match with the \dib9577 and \dib9632 DIBs, but conclusive identification of the weaker features has been difficult. \citet{gal17} were unable to confirm the presence of the weakest three C$_{60}^+$ bands in a sample of 19 heavily-reddened interstellar sightlines observed from the ground, and the studies of \citet{cor17} and \citet{lal18} concluded, at best, ambiguity regarding the presence of the \dib9428 band in HST and VLT spectra.

Although \citet{wal15,wal16} previously claimed interstellar detections of all five \cp\ bands, the presence of all the bands was not convincingly demonstrated in any single sightline \citep{cor17}. Reliable measurements of the three weaker \cp\ bands are particularly problematic because they fall in a wavelength region heavily obscured by telluric water vapor \citep[see][]{gal00,gal17}.  Telluric correction methods for weak interstellar absorption features are error prone \citep{lal18}, so to rigorously confirm the C$_{60}^+$ assignment, high signal-to-noise observations are required from outside the Earth's atmosphere. In the present study, we set out to obtain Hubble Space Telescope (HST) spectroscopy, unhindered by telluric absorption, to definitively confirm the presence (or absence) of the weaker \dib9348, \dib9365 and \dib9428 bands, which lie at the heart of the current C$_{60}^+$ debate. 

\section{Observations and data reduction}
\label{obs}

\begin{table}
\centering
\caption{Target star properties \label{tab:stars}}
\begin{tabular}{llcccc}
\hline\hline
Name&MK Type&$J$&\ebv&$v_{hel}$(K\,{\sc i})&S/N\\
&&(mag.)&(mag.)&(\kms)&\\
\hline
\multicolumn{6}{c}{Reddened Target Stars}\\
\hline
Cyg OB2\,\#5&O6+Of\,Ia&5.2&2.0&-10&600\\
HD\,195592&O9.7 Ia& 5.1 &1.1&-14& 500\\
BD+63\,1964&B0\,I&6.9&1.0&-23&700\\
HD\,169454&B1\,Ia&4.5&1.1&-9&700\\
HD\,190603&B1.5 Ia& 4.5 &0.7&-9& 500\\
HD\,136239&B2 Ia  & 5.7 &0.9&-15& 500\\
HD\,168625&B6 Iap & 5.1 &1.4&3& 500\\
\hline
\multicolumn{6}{c}{Unreddened Standard Stars}\\
\hline
$\tau$ Cma&O9\,III&4.7&0.1&---&1000\\
69 Cyg&B0\,Ib&6.1&0.1&---&800\\
HD\,36960&B1\,Ib&5.3&0.0&---&600\\
q Tau   &B6 IV  & 4.5 &  0.0 &---&700\\
\hline
\vspace{0mm}
\end{tabular}
\end{table}

Our method makes use of a novel Space Telescope Imaging Spectrograph (STIS) scanning technique to obtain considerably higher signal-to-noise (S/N) than usually achieved with this instrument. In addition to the the improved CCD illumination matching between science and flat-field exposures, which increases the overall efficacy of the flat fielding and fringe correction process, the benefits of STIS-scanning to obtain high-resolution HST spectra with unprecedented sensitivity were discussed in detail by \citet{cor17}. 

Observations of eleven stars were obtained during 2016 November to 2018 August as part of HST programs 14705, 15429 and 15478, including seven heavily-reddened stars and four lightly-reddened (or unreddened) spectral standard stars. Stellar types, $J$ magnitudes and reddenings (\ebv) are given in Table \ref{tab:stars}. Each star was observed in a single orbit using the STIS G750M grating with a central wavelength of 9336~\AA\ (covering the range 9050-9610~\AA, which unfortunately excluded the other strong, 9632~\AA\ \cp\ band). We used the $52''\times0.1''$ slit, with a plate scale of $0.05''$ per pixel and a spectral resolving power of $\sim10,000$. Following initial target acquisition and focusing maneuvers, a series of at least four exposures was obtained of each star, following the basic strategy of \citet{cor17}. For each exposure, the target star was positioned at CCD row 300, then scanned along the slit to row 1000 with the shutter open, resulting in a large portion of the CCD being exposed. The bottom 300 rows were avoided to reduce the detrimental effects of charge transfer inefficiency. Exposures were all performed in the `forward' scan direction to avoid the timing offsets in reverse scans identified by \citet{cor17}. For the removal of CCD fringing, a pair of tungsten flat field lamp exposures was obtained immediately after each pair of science target exposures; additional flats were obtained during Earth occultation to fill the remaining HST orbit time.  Pt/Cr-Ne arc lamp exposures were obtained at the beginning and end of each orbital visibility window, to correct for any dispersion drift in the STIS optics.

The basic data reduction and calibration procedures were described by \citet{cor17}. During spectral extraction, counts were integrated along each column, using a $3.5\sigma$ rejection threshold for any remaining bad pixels and cosmic ray hits.  Scattered light subtraction was performed with a low-order fit to the light under the occulting bar 2/3 of the way up the CCD.

\begin{figure*}
\centering
\includegraphics[width=0.95\textwidth]{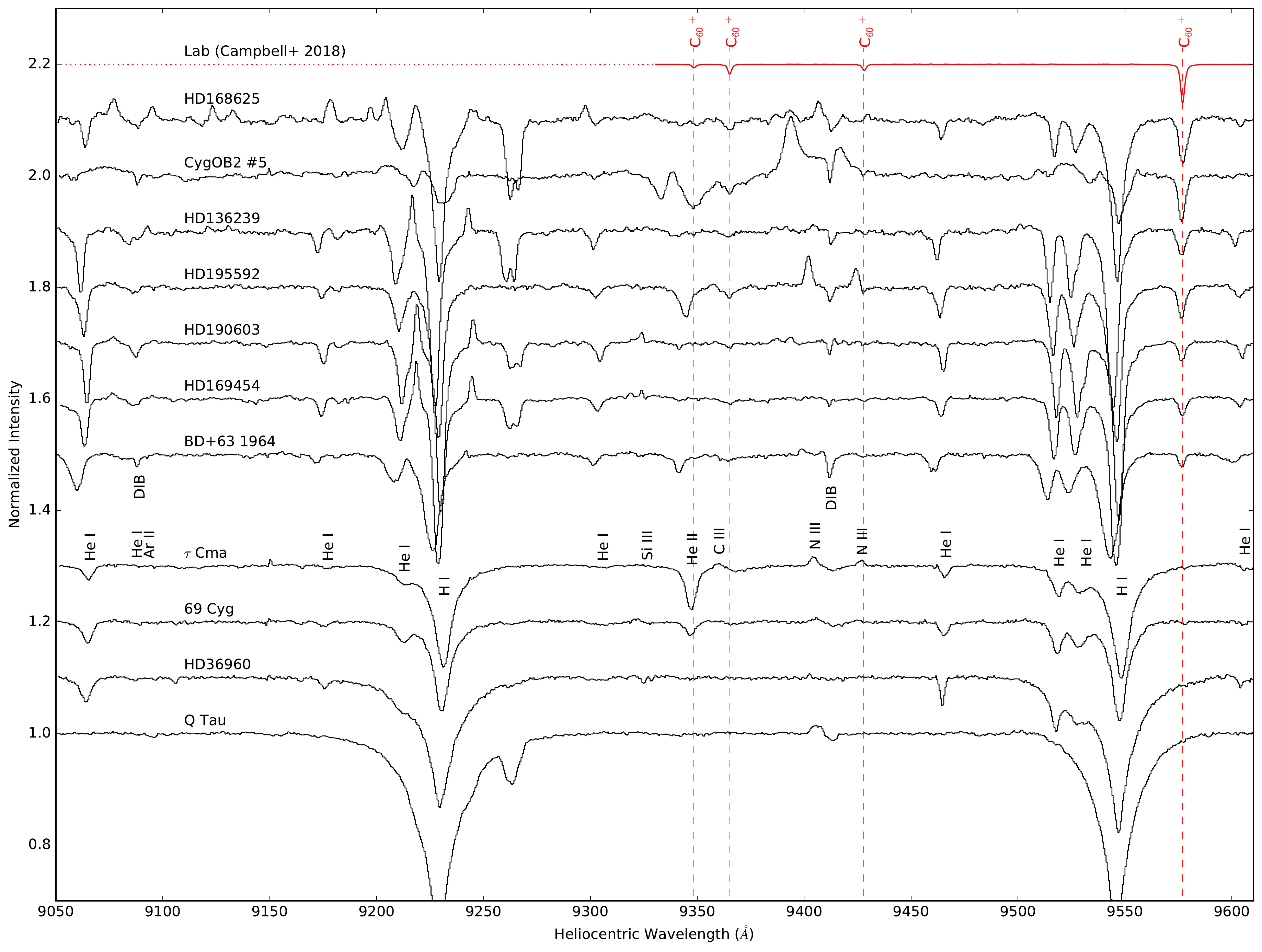}
\caption{Near-infrared HST STIS spectra of seven heavily-reddened interstellar sightlines and four unreddened standard stars (offset vertically for display). {A laboratory \cp\ spectrum from \citet{cam18} is shown for reference (see Section 3)}. Positions of the laboratory \cp\ bands, as well as known stellar lines and DIBs, are labeled.\label{fig:spectra}}
\end{figure*}

\section{Results}
\label{results}

The observed spectra are plotted in the heliocentric rest frame in Figure \ref{fig:spectra}. {A \cp\ reference spectrum is also shown, obtained from the high-resolution laboratory C$_{60}^+$--He measurements of \citet{cam18}. The laboratory cross sections were first corrected for an $\approx-0.6$~\AA\ wavelength shift caused by the `tagging' helium atoms \citep{cam16}, then scaled and converted to absorptivities, and finally convolved with a Gaussian line-broadening function of FWHM~=~30~\kms\ to match the STIS spectral resolution.}

The wavelengths of the 9577~\AA\ interstellar absorption bands attributed to \cp\ line up across all the reddened sightlines. This is not surprising because, as shown in Table \ref{tab:stars}, the interstellar \ki\ centroid velocities are relatively similar compared with the $\sim120$~\kms\ FWHM of this DIB. On the other hand, the stellar absorption features (some of the more prominent of which are labelled for $\tau$ Cma), show various Doppler shifts according to their respective stellar motions. We therefore confirm the presence of DIBs at 9088 and 9412~\AA\ in all our reddened sightlines (previously tentatively detected by \citealt{gal00} and \citealt{cor17}).

A feature corresponding to the 9365~\AA\ \cp\ band position can be seen in all the reddened sightlines, and a weaker feature at 9428~\AA\ is present in some of the sightlines. The relative band strengths and positions appear quite consistent with the laboratory \cp\ spectrum. 

\begin{figure*}
\includegraphics[width=\textwidth]{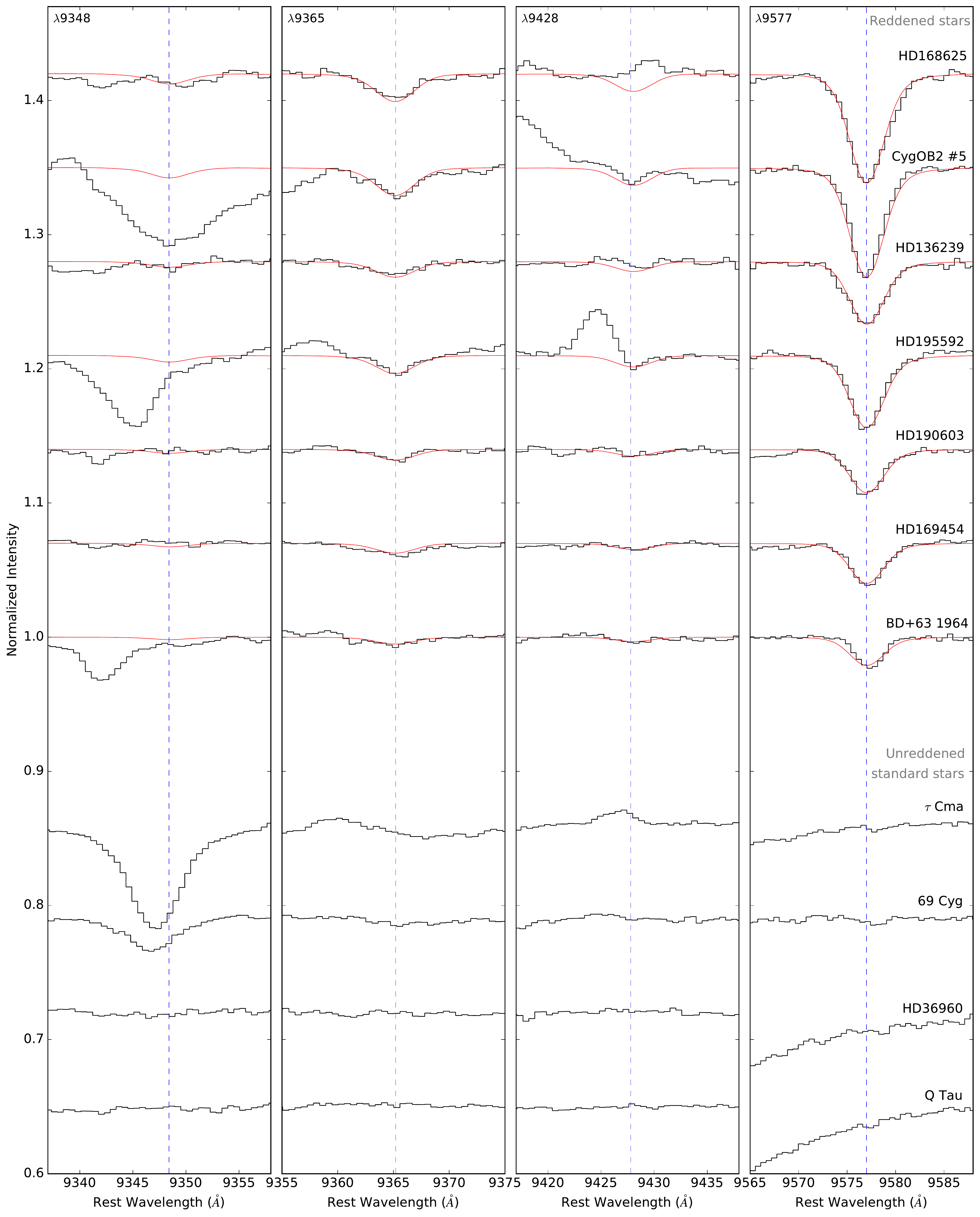}
\caption{STIS spectra of our program stars, zoomed in around the positions of the four observed \cp\ bands. Reddened spectra are in the interstellar K\,{\sc i} rest frame. {Red traces are \cp\ band predictions based on the laboratory spectrum of \citet{cam18}, broadened and scaled to match the \dib9577 observations}.\label{fig:zoom}}
\end{figure*}

A clearer (zoomed) view of the spectral regions of interest is shown in Figure \ref{fig:zoom}. {For each reddened star, a laboratory \cp\ spectrum has been overlaid, scaled to match the peak optical depth of the observed \dib9577 bands. It was necessary to apply an additional Gaussian broadening of FWHM~=~100~\kms\ to the lab spectrum, to obtain a reasonable agreement with the observed \dib9577 band widths (see also Section \ref{sec:discuss})}. In this Figure, the reddened target spectra have been shifted to the interstellar rest frame, defined by the centroid of the interstellar \ki\ \dib7698 absorption in each sightline. High resolution \ki\ echelle spectra were obtained for our target stars from archival data from the Keck, VLT, Mercator and CFHT telescopes. Keck HIRES \ki\ data (Cyg OB2\,\#5, HD\,168625 and HD\,190603) were described by \citet{cor07}, VLT spectra (HD\,136239 and HD\,169454) are from the ESO UVES science archive, Mercator HERMES observations (HD\,195592) are from H. Van Winckel (2019, private communication); see also \citet{ras11}, and the CFHT \espadons\ spectrum (BD+63\,1964) is from \citet{cor17}.

Identification of interstellar absorption features is facilitated by comparison of our target star spectra (upper traces in Figure \ref{fig:zoom}) with the unreddened standard star spectra (lower traces).  There is no convincing evidence for \cp\ absorption at 9348~\AA\ in any of our reddened sightlines. This is, however, consistent with the expected weakness of the \dib9348 \cp\ band, combined with the uncertainties in the observed stellar continuum due to the presence of noise and other overlapping stellar/interstellar features (in particular, a stellar He\,{\sc ii} line in the B0 and O-type stars). By contrast, the \cp\ \dib9365 band is clearly present, and closely matches the strength relative to \dib9577 found in the laboratory. There is evidence for the final, \dib9428 band towards Cyg OB2\,\#5, HD\,195592, HD\,190603, HD\,169454 and BD+63\,1964, but for HD\,136239 and HD\,168625, this band cannot be identified atop a fluctuating continuum. Determining the presence of the \dib9428 band is complicated by the nearby 9425~\AA\ N\,{\sc iii} emission line in the O-type stars, and as shown in Figure \ref{fig:spectra}, the HD\,168625 spectrum is contaminated by numerous emission features across the observed range, which preclude an accurate assessment of \dib9428 in this sightline.

\begin{table}
\centering
\caption{\cp\ band equivalent width measurements \label{tab:ew}}
\begin{tabular}{lclll}
\hline\hline
Sightline&FWHM$_{9577}$$^b$&$W_{9577}$&$W_{9428}$&$W_{9365}$\\
&(\AA)&(m\AA)&(m\AA)&(m\AA)\\
\hline
Cyg OB2\,\#5& 3.7& 327(5)& 20(5)& 72(5)$^a$\\
HD\,195592 & 4.1& 251(6)& 15(6)$^a$& 39(6)\\
BD+63\,1964& 2.8& 54(3)& 14(3)& 33(3)$^a$\\
HD\,169454 & 3.9& 132(3) & 13(3) &35(3) \\
HD\,190603 & 3.7& 134(5)& $<15$&26(5) \\
HD\,136239 & 4.3& 204(5)&$<15$ & 50(5)\\
HD\,168625 & 4.5&392(5)&$<15$$^a$ &68(5) \\[2mm]

Mean (all) & 4.1&214(2) &13(1)$^a$ &50(2) \\
Mean (early B)& 3.9& 138(2)&11(1)$^a$ &32(2)$^a$ \\

Campbell+ 2018$^c$ &1.7 & 138 & 21& 35\\
\hline
\end{tabular}
\parbox{0.9\columnwidth}
{\footnotesize $^a$Uncertain due to line blending or continuum uncertainties.\\
$^b$Band FWHM, deconvolved with respect to STIS spectral resolution.\\
$^c$Equivalent widths from the \citet{cam18} spectrum in Figure \ref{fig:means}b, normalized to the `mean (early B)' $W_{9577}$ value.
}
\end{table}

Equivalent widths ($W_{\lambda}$) were measured for the \dib9577, \dib9428 and \dib9365 bands (Table \ref{tab:ew}). Statistical ($1\sigma$) uncertainties (based on the measured RMS noise) are given in parentheses. When no \cp\ band was definitively detected, $3\sigma$ upper limits are given.

\begin{figure*}
\centering
\includegraphics[width=\textwidth]{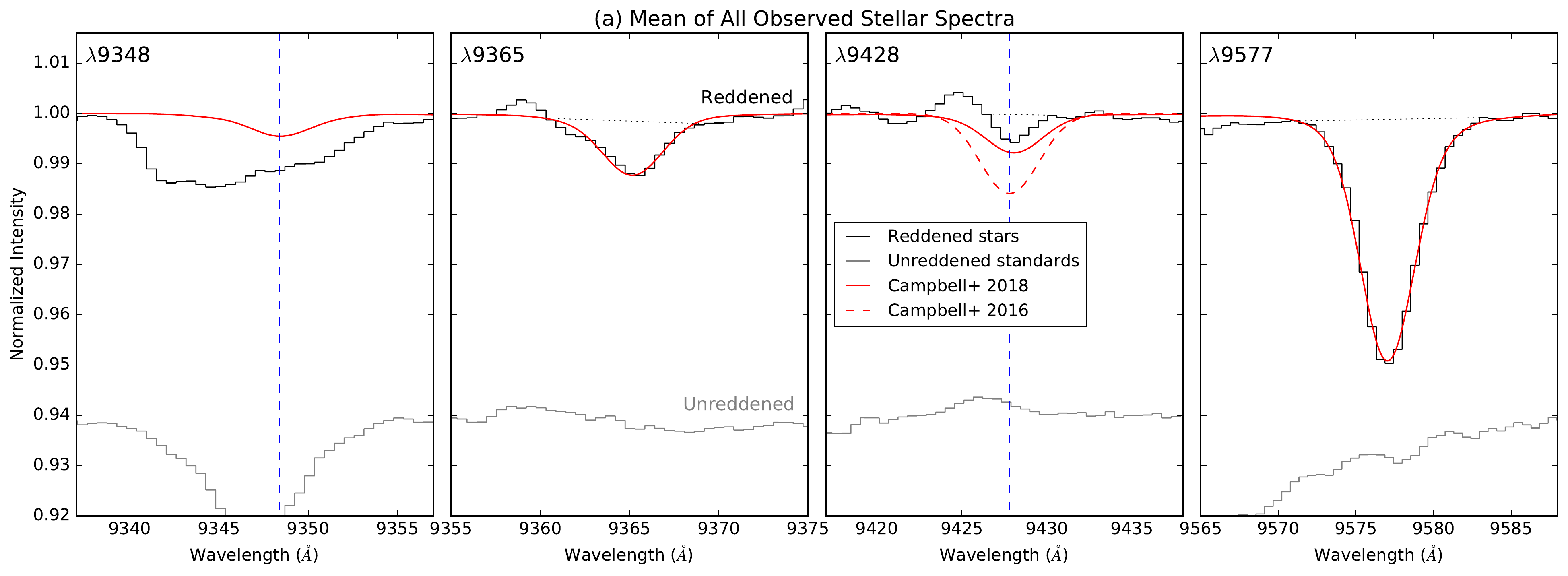}
\includegraphics[width=\textwidth]{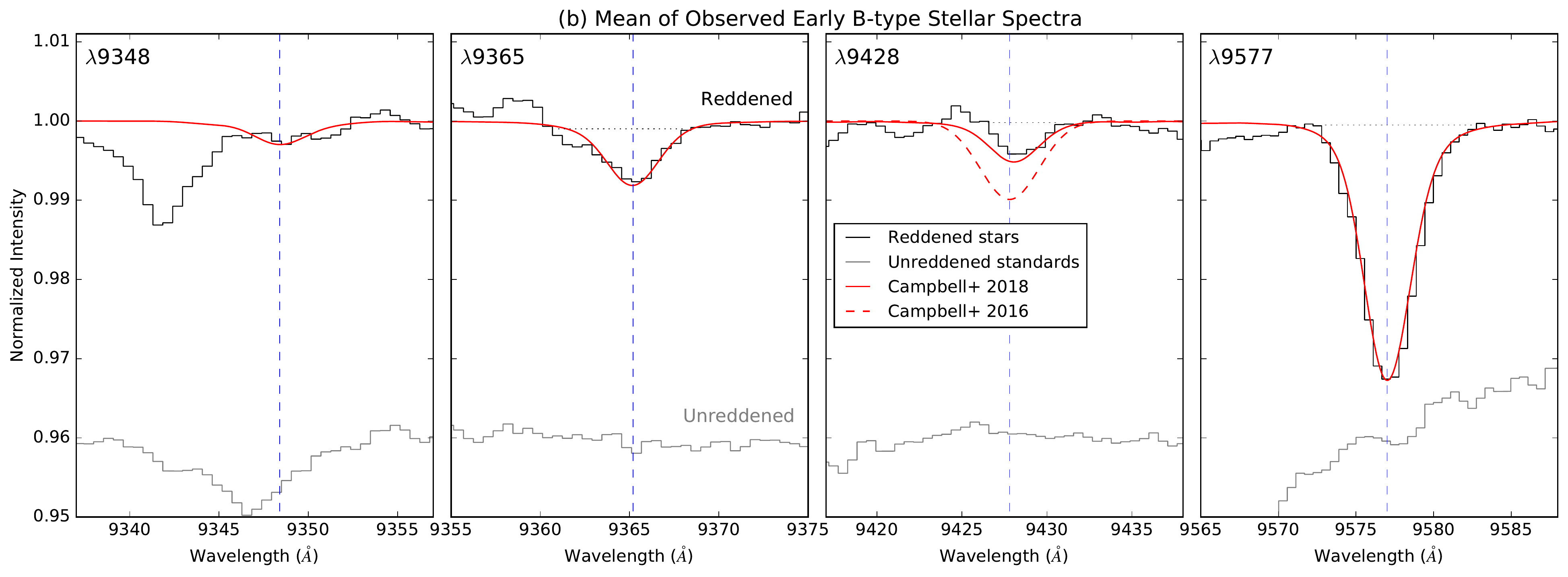}
\caption{Mean spectra for (a) all observed sightlines and (b) sightlines towards the early B-type stars. The corresponding mean spectra for the (unreddened) standards are shown in grey. Mean reddened spectra are in the interstellar K\,{\sc i} rest frame. Laboratory comparison spectra are overlaid in red, with a Gaussian broadening FWHM of 100~\kms\ in panel (a) and 80~\kms\ in panel (b). {Dotted black lines demark the integration areas used for equivalent width measurements.} \label{fig:means}}
\vspace*{0mm}
\end{figure*}

Averages of the continuum-normalized, Doppler-corrected spectral regions surrounding our observed \cp\ bands are shown in Figure \ref{fig:means}. Combining all the heavily-reddened sightlines helps reduce the statistical noise, averages out continuum uncertainties, and reduces the impact of individual stellar features due to their differing Doppler shifts between stars. The resulting mean spectra provide improved estimates for the interstellar \cp\ band strengths and profiles. The averages of the (unreddened) standard star spectra are also displayed. The same Doppler-broadened laboratory C$_{60}^+$ spectrum from Figure \ref{fig:zoom} is overlaid for comparison, scaled to the peak central depth of the \dib9577 band. For \dib9428, we also show a Gaussian comparison model based on the earlier \dib9577/\dib9428 band strength ratio obtained by \citet{cam16} (with a dashed line style).

As demonstrated in Figure \ref{fig:zoom} for Cyg OB2\,\#5, HD\,195592 and $\tau$~Cma, stellar line contamination of the weaker interstellar \cp\ bands is most severe for the observed O-type stars. We therefore generated an alternative mean spectrum for the reddened (and unreddened) stars, excluding the O-types. The B6\, Iap hypergiant (and LBV candidate) HD\,168625 was also excluded due to the large number of contaminating emission lines in its spectrum. The resulting `mean early B-type' spectrum is shown in Figure \ref{fig:means}b. The \dib9365 and \dib9577 bands provide a good match with the (scaled) laboratory spectrum. A weak absorption feature at 9362~\AA\ overlaps the \dib9365 band, and was identified as an interloping DIB by \citet{wal16}. An absorption feature at 9428~\AA\ is also clearly present in this spectrum, {apparently somewhat} weaker than its laboratory counterpart. Unfortunately, contamination from nearby stellar (or interstellar) lines hinders reliable measurement of the \dib9428 band strength. Rejection of the O-stars significantly improves the clarity of the continuum in the vicinity of the 9348~\AA\ \cp\ band, but no corresponding interstellar feature was detected. This is not surprising given the remaining uncertainties in the continuum, combined with the expected weakness of this band.

\section{Discussion}
\label{sec:discuss}

Our mean spectrum of reddened early B-type stars has a continuum S/N~$\sim1200$ and provides the clearest view to-date of the putative interstellar \cp\ absorption bands. Although the HST data are free from degradation due to telluric absorption, measurements of the three weaker \cp\ bands are still hindered by the presence of nearby stellar (or interstellar) lines, which introduce uncertainties into the continuum level and band profiles. Nevertheless, we confidently determined the presence of absorption features at interstellar rest wavelengths of $9365.1\pm0.1$ and $9428.5\pm0.2$, $9577.1\pm0.1$~\AA. These match closely the laboratory wavelengths of [$9365.2\pm0.2$, $9427.8\pm0.2$, $9577.0\pm0.2$~\AA] determined by \citet{cam16}, and are also in reasonable agreement with the wavelengths of [$9364.8\pm0.1$, $9427.6\pm0.2$, $9576.5\pm0.2$~\AA] and [$9364.9\pm0.1$, $9427.5\pm0.1$ and $9576.7\pm0.1$~\AA] from the independent \cp\ helium droplet experiments of \citet{kuh16} and \citet{spi17}, respectively.

The \dib9428 band has the largest wavelength discrepancy, with an offset of between 0.5-1.3~\AA\ (1-2 pixels) with respect to the range of laboratory measurements. Such a relatively small wavelength shift could be the result of contamination of the HST spectra by interloping stellar or interstellar features that skew the observed band profile. Inaccuracy in the laboratory wavelength is also possible as a result of uncertainty in extrapolating the measured C$_{60}^+$--He$_n$ wavelengths down to $n=0$. Small (sub-Angstrom) wavelength shifts may also be introduced as a result of $^{13}$C-substitutions \citep{cam18}. To rule out other observational errors, we searched our spectra for CCD defects such as residual bad pixels and fringing artifacts, but no repeatable artifacts were found across all our observations. 

High-resolution studies of interstellar K\,{\sc i}, H\,{\sc i}, Ca\,{\sc ii} and Ti\,{\sc ii} absorption \citep[\eg][]{wel10,cox06} show that the differing ionisation and depletion levels of these gases can lead to significant differences between their respective radial velocity profiles. Differences between the \ki\ and \cp\ mean radial velocities may therefore be expected, which would introduce errors into our measured \cp\ band rest wavelengths. However, such differences are likely to be small compared with our $\sim30$~\kms\ spectral resolution. Furthermore, any pseudo-random differences between the \ki\ and \cp\ line-of-sight velocity distributions would tend to be averaged out in our mean spectra.  Adopting a generous uncertainty of $\pm10$~\kms\ on the \cp\ rest velocity translates to only $\pm0.3$~\AA\ on the measured wavelength, which does not significantly alter the conclusions of our study.

The interstellar \dib9577 band shape is well described by a Gaussian profile in all sightlines apart from HD195592, which has more extended, Lorentzian-type wings. The mean observed profile has a slightly enhanced red wing, which is qualitatively similar to the laboratory profile of C$_{60}^+$--He from \citet{cam18}. The \dib9577 band FWHM measurements are given in Table \ref{tab:ew}, with an average value of $4.2\pm0.1$~\AA. This is relatively broad compared with the ($\approx0.95$~\AA) spectral resolution, resulting in a deconvolved mean FWHM of 4.1~\AA, which is more than a factor of two broader than the laboratory band FWHM of 1.7~\AA. Assuming \dib9577 is due to a rovibronic transition of \cp, the broad profile can be explained by an (unresolved) manifold of rotational lines, that gives rise to a rotational contour, the width of which varies as a function of rotational temperature \citep[see][]{edw93}. In our case, the observed FWHM~$\approx4$~\AA\ suggests a mean \cp\ rotational temperature $\sim100$~K in the diffuse ISM \citep[see also][]{foi97,cam18}.

As shown by Figure \ref{fig:means}, the degree of broadening required to fit the \dib9365 band is similar to that of \dib9577, suggesting a similar rotational contour. This is consistent with the conclusion of \citet{lyk19}, that the \dib9365 and \dib9577 bands arise in the same $^2A_u\longrightarrow^2A_g$ electronic transition from the $v=0$ ground vibrational state of \cp, with $v'=0$ in the excited state for \dib9577 and $v'=1$ for \dib9365. By analogy, the \dib9428 band is believed to originate from the $^2A_u\longrightarrow^2B_g$ partner transition, to the Jahn-Teller-split excited state (with $v'=1$). Although the \dib9428 profile is difficult to reliably measure in individual sightlines (Figure \ref{fig:zoom}), our mean spectra (Figure \ref{fig:means}) show that it may be narrower than the \dib9365 and \dib9577 bands, but this result is uncertain due to possible spectral contamination.

Our observed ratio of equivalent widths $W_{9577}$:$W_{9428}$:$W_{9365}$=1.0:0.08:0.23 in the mean early B-type spectrum is comparable with the ratios of 1.0:0.15:0.25 derived from the convolved C$_{60}^+$--He laboratory spectrum in Figure \ref{fig:means}. The latter values differ from those published by \citet{cam18} because they represent band-integrated equivalent width ratios rather than the peak cross-section ratios given in their study. Whereas the observed \dib9365/\dib9577 equivalent width ratio of 0.23 matches closely with the laboratory ratio of 0.25, a relative weakness is apparent for the \dib9428 band (with \dib9428/\dib9577 = 0.08 \emph{vs.} 0.15) --- the apparent weakness is even more pronounced when compared with the earlier cross section ratio of \dib9428/\dib9577 = 0.3 from \citet{cam16}, and may explain why this band was not detected in the ISM by \citet{gal17,gal17b}. Such a discrepancy could result from subtle perturbations to the transition strengths due to the attached helium atom, or from differing physical conditions between the laboratory and ISM. For example, \citet{lyk19} suggested that the relative \cp\ band strengths could vary as a result of temperature-dependent rotation-vibration coupling, that would directly alter the intensities of the \cp\ rovibronic transitions.

\section{Conclusion}
\label{conclusion}

We have obtained high-S/N HST spectra of the \dib9348, \dib9365, \dib9428 and \dib9577 \cp\ bands along seven heavily reddened interstellar lines of sight. The strong \dib9577 band and weaker \dib9365 band are clearly identified in all sightlines, with wavelengths and equivalent width ratios closely matching the latest laboratory data. A weak \dib9428 band can also be seen in the early B-type sightlines where contamination from stellar features is less severe. The \dib9428 wavelength and profile appear to differ from those expected based on the laboratory measurements of \citet{cam18}, and its {measured equivalent width} is $\sim50$\% less than expected, but residual stellar/interstellar contamination cannot be ruled out as a possible explanation for these small discrepancies. The \dib9348 band could not be detected due to its intrinsic weakness and overlapping stellar lines. In summary, we confirm the presence of all three expected \cp\ bands in the diffuse interstellar medium, with strength ratios consistent with those measured in the laboratory for C$_{60}^+$--He at very low temperature. We consider this the first robust detection of the \dib9428 interstellar band. Combined with prior, ground-based observations of the \dib9365, \dib9577 and \dib9632 bands \citep[\eg][]{wal15,wal16,gal17b,lal18} our HST spectra place the detection of interstellar \cp\ beyond reasonable doubt.

The confirmation of interstellar \cp\ represents a breakthrough in our understanding of chemical complexity in the diffuse ISM, dramatically increasing the size limit for known carbon-bearing molecules in low-density, strongly-irradiated environments, and bringing a new understanding of the types of molecules that may be responsible for the remaining (unidentified) DIBs. Further high-sensitivity observations are recommended to better constrain the strengths and profiles of the weaker \cp\ bands, combined with additional laboratory and theoretical studies that may enable the exploitation of the \cp\ bands as probes of interstellar physics and chemistry.

\acknowledgments
This work is based on observations made with the NASA/ESA Hubble Space Telescope, obtained at the Space Telescope Science Institute, which is operated by the Association of Universities for Research in Astronomy, Inc., under NASA contract NAS 5-26555. P.J.S. thanks the Leverhulme Trust for an Emeritus Fellowship award. F.N. was supported through Spanish grants ESP2015-65597-C4-1-R and ESP2017-86582-C4-1-R.


\end{document}